\documentclass[a4paper]{article}

\usepackage{INTERSPEECH2022}

\usepackage{caption}
\usepackage{subcaption}

\title{A Simple Feature Method for Prosody Rhythm Comparison}
\name{Mariana Julião$^{1,2}$, Alberto Abad$^{1,2}$, Helena Moniz$^{1,3}$} 
\address{
  $^1$INESC-ID, Lisbon, Portugal\\
  $^2$Instituto Superior Técnico, University of Lisbon, Portugal\\
  $^3$FLUL - School of Arts and Humanities, University of Lisbon, Portugal}
\email{mjuliao@hlt.inesc-id.pt, alberto.abad@inesc-id.pt, helena.moniz@inesc-id.pt}

\begin{document}

\maketitle
\begin{abstract}
Of all components of Prosody, Rhythm has been regarded as the hardest to address, as it is utterly linked to Pitch and Intensity. Nevertheless, Rhythm is a very good indicator of a speaker's fluency in a foreign language or even of some diseases. Canonical ways to measure Rhythm, such as $\Delta C$ or $\%V$, involve a cumbersome process of segment alignment, often leading to modest and questionable results. Perceptively, however, rhythm does not sound as difficult, as humans can grasp it even when the text is not fully intelligible. In this work, we develop an empirical and unsupervised method of rhythm assessment, which does not rely on the content. We have created a fixed-length representation of each utterance, Peak Embedding (PE), which codifies the proportional distance between peaks of the chosen Low-Level Descriptors. 
Clustering pairs of small sentence-like units, we have attained averages of 0.444 for Silhouette Coefficient using PE with Loudness, and 0.979 for Global Separability Index with a combination of PE with Pitch and Loudness. Clustering same-structure words,  we have attained averages of 0.196 for Silhouette Coefficient and 0.864 for Global Separability Index for PE with Loudness.
 
\end{abstract}
\noindent\textbf{Index Terms}: prosody, rhythm embeddings, rhythm comparison, prosody assessment

\section{Introduction}

The word Rhythm comes from a Greek word 
which means regular movement, as the pace of a poem or the metrics of a verse. Prosody-wise, Rhythm is the cadence and alternation of stressed and unstressed syllables. The components of prosody -- Pitch, Loudness, Rhythm -- are often referred to as if they were independent and disentagleable. That is true for Pitch and Loudness, which are the perceptual qualities of physical quantities of the sound -- F0 and intensity, respectively -- which can have well-defined values in the spectrum. Rhythm, however, is not a physical quantity that is measurable on a frame-basis. Instead, it is a measure of how the other two behave in time. 

The concept of Rhythm is often erroneously taken as a synonym of speech rate~\cite{dellwo2003relations, wagner2004introducing}. Whereas Rhythm is about the proportion of the distances between stressed syllables, speech rate is about how fast a speaker utters a sentence-like unit (SLU). Consider, for example, the sentence "Please call Stella" from the VCTK corpus\footnote{More on this corpus later on.}. The distance between the first and the second stressed syllables and the distance between the second and the third stressed syllables approximately keep their proportion across different speakers. However, the length of this utterance ranges from 0.9 seconds (for fast speakers) to 1.71 seconds (for slow speakers). Considering these extreme cases, we can say that the Rhythm is the same, but the speech rate is very different.

Most of the features that have been used to describe Rhythm do not encompass how the stresses are distributed along the utterances, but consist of average metrics instead.
In this work, we look for a feature that allows to compare prosody while encompassing Rhythm information, and that allows for a direct and simple comparison of two SLUs. This is mostly attainable if all SLUs are represented by feature vectors of the same length. In a previous work~\cite{juliao2022can}, we explored the use of prosody transfer embeddings for prosody comparison. 
Here, we propose to extract a vector of fixed-length enconding which represents the maxima of prosodic features in their position in the segment. To this we call Peak Embedding (PE).
We take 10 SLUs which are phonetically rich and rhythmically contrastive, utterred by many speakers, and evaluate their clustering. Assuming that SLUs with the same text share Rhythm, Intonation and Energy to a large degree, we hypothesise that these features are as good to discriminate prosody as they are able to generate good clustering metrics. Based on the same principles, we take 5 words with two different rhythmic structures, and evaluate the extent to which these two rhythmic classes are clustered.

\section{Related Work}~\label{s:rw}

To account for the rhythmic distinction between stressed-timed and syllable-timed languages, and clarify how Rhythm might be extracted from the speech signal, Ramus~\cite{ramus1999correlates} proposed three different features, each taking one value per utterance: (1) the proportion of vocalic intervals within the sentence (\%V), (2) the standard deviation of the duration of vocalic intervals within each sentence ($\Delta V$), as well as  (3) the standard deviation of the duration of consonantal intervals within each sentence ($\Delta$C). 
Other metrics have also been proposed, such as PVI~\cite{grabe2002durational}, based on the sum of the differences between pairs of consecutive intervals, or YARD~\cite{wagner2004introducing}, a metric similar to PVI in which z-transformed syllable durations are used for calculation, rather than raw intervals.

In a large review of Rhythm metrics, Arvaniti~\cite{arvaniti2012usefulness} concluded that in spite of their abundant use, most of these metrics are largely affected by methodological choices like the type of speech (spontaneous/read), or the syllable complexity of the texts. As Wagner~\cite{wagner2004introducing} states, Ramus' Rhythm measures ``provide a working discrimination of rhythmic classes based on syllable complexity, but they miss a representation of the sequential nature of Rhythm''. Furthermore, this class of metrics requires a considerable overhead of segment alignment and processing.

Other methods rely on the Envelope Modulation Spectrum, proposed by Tilsen~\cite{tilsen2008low}, which uses spectral analysis of the amplitude envelope of vocalic energy. Studies on intelligibility for different diseases or health impairments have used this method later on~\cite{liss2010discriminating, leong2014impaired, vaysse2021automatic}.

In the speech conversion realm, \cite{qian2020unsupervised} developed a system to convert the Pitch, Content, Rhythm or Timber from a source sentence into the same component of a target utterance. This was achieved by using a triple information bottleneck on a system with an encoder for each component (except timber). MOS results showed that not only the conversion of Rhythm was less successful than the conversion of Content and Pitch, but also that when Rhythm was paired with another component the results were worse than with the other component alone. This points towards the difficulty in actually separating Rhythm from the other components. More recently, \cite{qian2021global} developed another system of speech style conversion, this time focusing on Rhythm. Interestingly, the MOS were better than in~\cite{qian2020unsupervised} and Rhythm was not assessed separately from the other components.

\section{Methodology}


The methodology we follow consists of analysing data corresponding to different sentence-like units (SLU) with Peak Embeddings based on two different features: Pitch and Loudness. As a baseline, we used temporal functionals.
We take as a premise that (1) two SLUs with the same content and context, and uttered by native speakers are very similar prosody-wise; (2) a feature is as good to encode prosody as it is able to lead to well-separated clusters of contrastive utterances. 
At second step, we inspect the clustering behaviour of words, assuming that words with the same stressed syllable distribution and intonation have similar rhythmic patters. We analyse this by using specific metrics to assess clustering. We also inspect the t-SNE projections~\cite{van2008visualizing}\footnote{t-SNE is a method that permits visualization of data in a two-dimensional space.} of the utterances.

\subsection{Features}


\subsubsection{Baseline Functionals}~\label{ss:bf}

The Geneva Minimalistic Acoustic Parameter Set (eGeMAPS)~\cite{eyben2015geneva} is a set of functionals computed on top of LLDs, specifically chosen to provide a common ground for research in various areas of automatic voice analysis, namely paralinguistics. For a baseline, we used the six temporal features of eGeMAPS: rate of loudness peaks, mean length and standard deviation of continuously voiced regions, mean length and standard deviation of unvoiced regions, number of continuous voiced regions per second (pseudo-syllable rate). Each of these features corresponds to one global value per SLU, which means that, in practice, it is not possible to reflect the local properties of the utterance, on which Rhythm heavily relies. Having these features as a baseline shall help us better understand the contribution of the Peak Embeddings we are proposing here. Also, these features resemble the features of~\cite{ramus1999correlates}, described in Section~\ref{s:rw}, thus providing us with an interesting reference.

\subsubsection{Peak Embeddings}~\label{ss:pe}

The Peak Embeddings (PE) we propose are extracted as follows: (1) Take a low-level descriptor of a speech segment; (2) Normalise; (3) Smooth with moving average; (4) Divide into 10 equal-sized chunks; (5) Do max-pooling of each chunk.
As a result, these embeddings have a fixed dimension, which allows them to store information about the proportional regions where peaks occur, thus also encompassing durational aspects. We have extracted PE for Pitch and Loudness, as these are the components of Prosody with direct acoustic correlates. By using these embeddings we are incorporating the temporal dimension of Pitch and Loudness. For Pitch and Loudness, we used the LLDs of eGeMAPS. We have also combined the PE of Pitch with the PE of Loudness, both by concatenation (;) and sum (+).  




\subsection{Clustering Metrics}

\subsubsection{Silhouette Coefficient}

The Silhouette Coefficient~\cite{rousseeuw1987silhouettes} measures how the samples in a data set are similar to the ones of its own class when compared to samples in other classes. It ranges from -1 to 1, where negative values mean assignments to the wrong cluster, 0 indicates cluster overlap, and 1 indicates the best separation possible. 
For each sample x, the Silhouette Score is 
\begin{equation}
    S = \frac{b(x)-a(x)}{max(a(x),b(x))} \, ,
\end{equation}
where $a(x)$ is the mean intra-cluster distance and $b(x)$ is mean nearest-cluster distance. The Silhouette Coefficient is the mean Silhouette Score of all samples. 

\subsubsection{Geometrical Separability Index}

Thornton Index (TH)~\cite{thornton1998separability}, also known as Geometrical Separability Index (GSI), corresponds to the fraction of a set of points that have the same classification labels as their first neighbour. It can be described as
\begin{equation}
    GSI = \frac{1}{n}\Sigma^n_{i=1}f(x_i,x'_i) \, ,
\end{equation}
where $x'_i$ is the nearest neighbour of $x_i$, $n$ is the number of points, $f$ is a function that is 1 if $x_i$ and $x'_i$ belong to the same class and zero otherwise. Therefore, GSI will tend to 1 if opposite labels exist in well-separated groups~\cite{acevedo2019measuring}. 


\section{Experiments}

\subsection{Corpora}

\subsubsection{VCTK Data Set}

We have used the CSTR VCTK Corpus~\cite{veaux2016superseded}. It comprises 110 English speakers of different varieties of English, with approximately 400 utterances per speaker. These correspond to a phonetically-rich paragraph, which is common to all speakers, and a newspaper text which is different for all speakers. We considered only the sentences of the phonetically rich paragraph. For more size homogeneity, we cropped the original sentences into the resulting SLU in Table~\ref{tab:slu}.

\begin{table}[t]
  \caption{Sentence-Like Units. $\sim$ means that a part of the original utterance was omitted on that place.}
  \label{tab:slu}
  \centering
  \begin{tabular}{ll}
    \toprule
    \textbf{ID}              & \textbf{Text}                \\
    \midrule
001 & Please call Stella.                             \\
002 & Ask her to bring these things $\sim$                  \\
003 & Six spoons of fresh snow peas $\sim$                  \\
004 & $\sim$ a big toy frog for the kids.                    \\
005 & She can scoop these things into three red bags $\sim$ \\
006 & $\sim$ they act like a prism and form a rainbow.       \\
007 & The rainbow is a division of white light $\sim$       \\
008 & These take the shape of a long round arch $\sim$      \\
009 & $\sim$ a boiling pot of gold at one end.              \\
010 & People look but no one ever finds it.      \\
    \bottomrule
  \end{tabular}
\end{table}

\subsubsection{Intonation Data Set}

We used a small intonation data set, comprising 20 original stimuli recorded by a native female speaker of Standard European Portuguese and reproductions of it by 17 different speakers. 
The 20 original stimuli correspond to the possible combinations of five words: \textit{Banana}, \textit{Bolo}, \textit{Gelado}, \textit{Leite}, and \textit{Ovo}\footnote{The translation of these words in English is, respectively, \textit{Banana}, \textit{Cake}, \textit{Ice Cream}, \textit{Milk}, and \textit{Egg}.} with four different intonations: \textit{Declarative}, \textit{Interrogative}
, \textit{Pleasure}, and \textit{Displeasure}. Each utterance corresponds to one word uttered with one particular intonation. In total, there are 340 imitation utterances, of which 320 are labelled as good, which are the ones we used.


\subsection{Experimental Set Up}

\subsubsection{Sentence-Like Units}

We computed the clustering metrics of all 45 ($C_{10,2}=45$) possible pairs of SLUs, for all features we are considering. To better understand the utility of each feature, we computed the average value of each cluster metric for each feature and all pairs (Table~\ref{tab:cm}).

\subsubsection{Isolated Words}

We computed the clustering metrics of all 4 intonation types, for all features we are considering. We compared the Rhythm of different words by assuming that words with a similar syllabic structure have similar Rhythm. Given five words with stress on the penultimate syllable (paroxytone), we split them into two classes: words with 3 syllables (``ba\textbf{na}na'', ``ge\textbf{la}do'') and words with 2 syllables (``\textbf{bo}lo'',``\textbf{lei}te'', ``\textbf{o}vo''). We are comparing within each intonation type, as we expect that the same word can have different rhythms for different intonations. Average results for each intonation type are on the right-most columns of (Table~\ref{tab:cm}).

\subsection{Results}

\begin{figure*}[h]
\centering

\subfloat{\includegraphics[width = 0.4\textwidth]{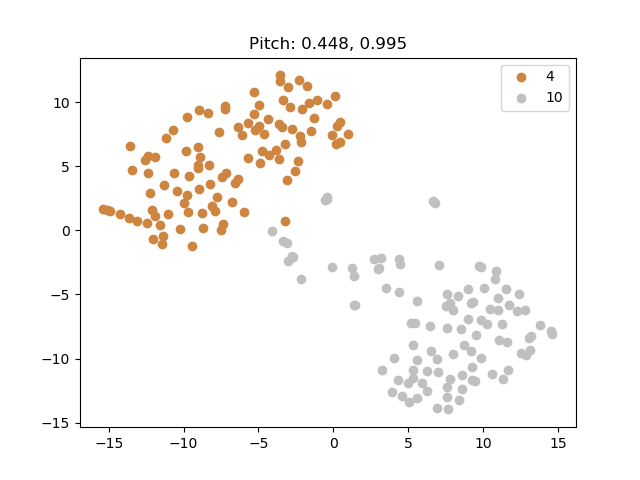}} 
\subfloat{\includegraphics[width = 0.4\textwidth]{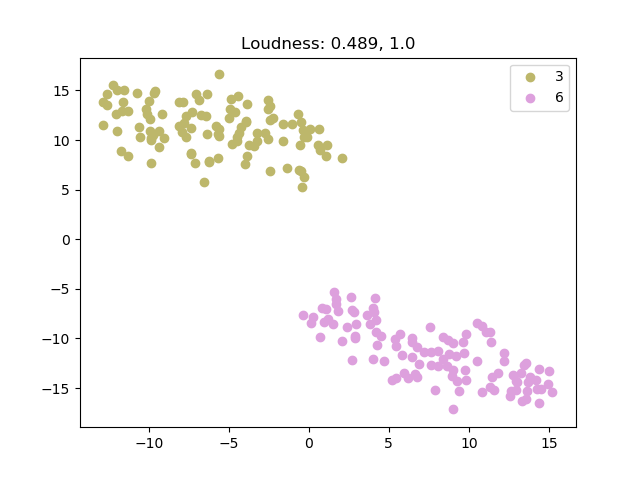}}\\
\subfloat{\includegraphics[width = 0.4\textwidth]{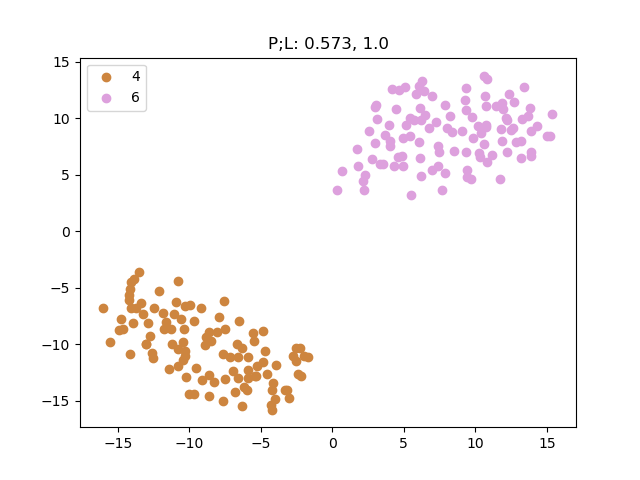}}
\subfloat{\includegraphics[width = 0.4\textwidth]{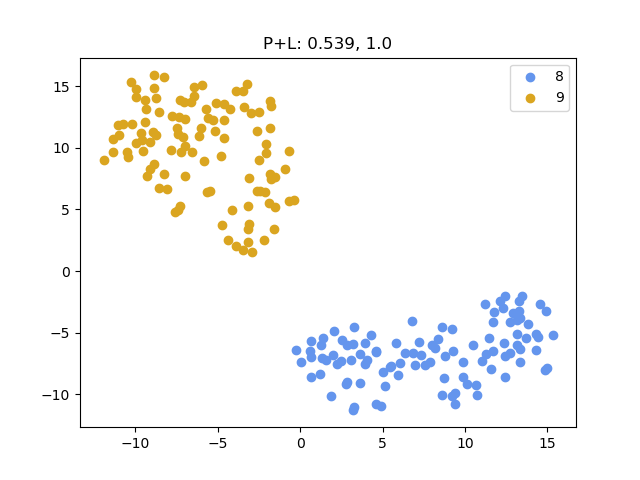}}

\caption{t-SNE projections of the pairs of SLUs which attained the best clustering results for Peak Embeddings on based on Pitch (P), Loudness(L), Pitch;Loudness (P;L) and Pitch+Loudness (P+L). Values on top of the plots correspond to SC and GSI, respectively).}
\label{fig:tsne}
\end{figure*}

\begin{figure}[t]
  \centering
  \includegraphics[width=\linewidth]{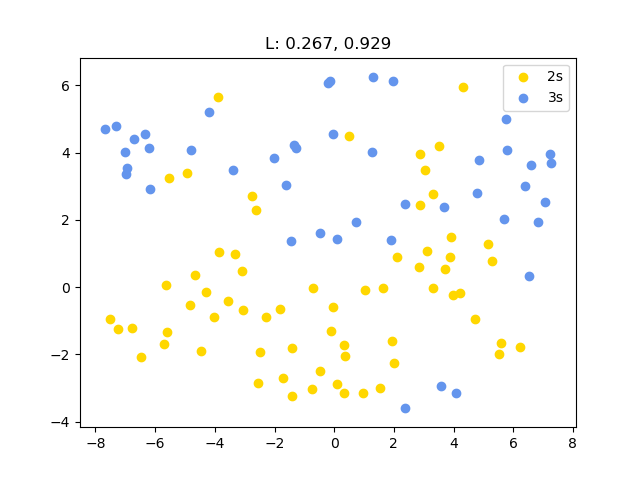}
  \caption{t-SNE projection for words of different syllable structure, with intonation \textit{Displeasure}, Peak Embeddings of Loudness.}
  \label{fig:tsnebnn}
\end{figure}

The average Clustering Metrics are in Table~\ref{tab:cm}. 
The best clustered SLU pairs with different Peak Embeddings are in Table~\ref{tab:bwslu}. The t-SNE projections of the pairs of SLUs with the best clustering results for different PEs are in Figure~\ref{fig:tsne}. The t-SNE projection of the intonation type with the best clustering results for words of different Rhythm is in Figure~\ref{fig:tsnebnn}.


\begin{table}[t]
  \caption{Clustering Metrics: average for the 45 pairs of SLU; average for the 4 intonation types.}
  \label{tab:cm}
  \centering
\begin{tabular}{lllll}
    \toprule
                     & \multicolumn{2}{c}{\textbf{SLU}}      & \multicolumn{2}{c}{\textbf{Word}} \\
                     \midrule
                     & SC  $\uparrow$            & GSI   $\uparrow$     & SC   $\uparrow$ 
                     & GSI   $\uparrow$     \\
                       \midrule
Baseline Funcs.      & 0.147           & 0.730      & 0.112       & 0.796      \\
    \midrule
    \midrule
Pitch                & 0.371           & 0.951      & 0.040       & 0.744      \\
Loudness             & \textbf{0.444}           & 0.958      & \textbf{0.196}       & \textbf{0.864}      \\
Pitch; Loudness      & 0.394           & \textbf{0.979}      & 0.137       & 0.841      \\
Pitch+Loudness       & 0.400           & 0.955      & 0.127       & 0.795     \\
    \bottomrule
  \end{tabular}
\end{table}


\begin{table}[t]
  \caption{SLUs with best(+) and worst(-) clustering metrics for each feature type. (Good results are easier to attain when SLUs are prosodically more different.)}
  \label{tab:bwslu}
  \centering
  \begingroup
  \begin{tabular}{c|ccc|ccc}
    \toprule
  & \multicolumn{3}{c}{\textbf{Pitch}} & \multicolumn{3}{c}{\textbf{Loudness} }\\
           \midrule
+ & (4,10)  & (3,9)  & (8,9)  & (3,6)    & (3,9)   & (5,10)  \\
 & (5,9)   & (6,10) & (5,10) & (6,10)   & (4,10)  & (5,9)   \\
          \midrule
- & (5,6)   & (7,9)  & (2,8)  & (2,7)    & (1,5)   & (3,8)   \\
 & (5,8)   & (3,8)  & (9,10) & (1,7)    & (1,9)   & (7,9)   \\
          \midrule
 & \multicolumn{3}{c}{\textbf{P;L}}   & \multicolumn{3}{c}{\textbf{P+L}}      \\
          \midrule
+ & (4,6)   & (6,8)  & (6,10) & (8,9)    & (3,8)   & (1,9)   \\
 & (5,10)  & (5,9)  & (3,9)  & (4,10)   & (3,6)   & (5,9)   \\
          \midrule
- & (1,9)   & (7,9)  & (1,7)  & (5,6)    & (3,8)   & (1,9)   \\
 & (5,8)   & (1,2)  & (2,8)  & (1,6)    & (7,9)   & (1,5)  \\
    \bottomrule
  \end{tabular}
  \endgroup
\end{table}







\section{Discussion}

On the left-most columns (SLU) of Table~\ref{tab:cm} we see that Peak Embeddings based on Loudness achieve the best clustering results of all for SC, and the second best for GSI. This indicates that a temporal representation of Loudness can account for most of the prosodic differences between SLUs. Cluster compactness (measured by the SC) does not improve when we combine Pitch to Loudness. Global Separability improves when we concatenate Loudness and Pitch; it does not improve when we add them. This means that Loudness and Pitch provide different types of informativeness and that their peaks do not necessarily coincide. 
On the right-most columns (Word) of Table~\ref{tab:cm} we see that Peak Embeddings based on Loudness achieve the best clustering results of all for both metrics. For single words, the information of Loudness in time is what better accounts for a word's rhythm. We see in the example of Figure~\ref{fig:tsnebnn} that, although with little compactness, words with two syllables tend to be separated from words of three syllables.
However, the most relevant result in Table~\ref{tab:cm} is that in all cases Peak Embeddings attain better results than the Baseline Results, which means that Rhythm comparison based on functionals is clearly suboptimal. 
Percentages of consonants and vowels do not give us much information on the rhythm of a specific sentence, as it does not have information on the places where stresses (peaks) happen. 

Looking at SLUs with the best and worst separation, in Table~\ref{tab:bwslu}, some sentence pairs are noteworthy. For all features, pairs (7,9) and (1,9), are amongst the pairs with worst separability. As these sentences are not particularly similar in terms of prosody, we argue this may be due to prosody caracteristics that are more difficult to encode or even high speaker variability. (5,9) is amongst the best for all the considered features; (3,9), (5,10), (4,10), (6,10) are amongst the best for three of the features. They share the fact that in each pair none of the sentences is trimmed at the same end: either they are trimmed on opposite ends or one is trimmed and the other is not. (8,9) is only at the top six for Pitch and P+L, possibly because Pitch is what better accounts for the different intonations of the SLUs (rising vs falling).

In any case, we ought to say that the results we have for separability are fairly high, as we saw in Table~\ref{tab:cm}. The difference between most pairs' clustering is more about the compactness of the cluster than it is about the separability of the set. This means that these results should be taken with some caution, as the metrics for the SLU pairs are very close and the threshold we are imposing (six best SLUs) can make us ignore relevant pairs.

\section{Conclusion}

In this work, we have assessed a method for representing maxima of prosody Low-Level Descriptors in time, Peak Embeddings. We have explored how they can account for the rhythmic properties of variable-length speech segments, by assessing how different pairs of SLUs and words were clustered. For SLUs, we reached an average GSI of 0.979 when concatenating the Peak Embeddings of Pitch and Loudness, and an average of 0.444 SC for Peak Embeddings of Loudness only, against an SC of 0.147 and a GSI of 0.730 for the temporal features of eGeMAPS, which we took as baseline. For words, we reached an average GSI of 0.864 and 0.196 SC with for Peak Embeddings of Loudness, against an SC of 0.112 and a GSI of 0.796 for baseline features.
This strengthens our conviction that a description of Rhythm -- or Prosody, in general -- without local sensitivity is doomed to be poorly informative, and that a temporal representation of prosodic features is very relevant. As future work, we would like to investigate how robust this method is across languages and utterance types.

\section{Acknowledgements}

This work was supported by national funds through Funda\c c\~ao para a Ci\^{e}ncia e a Tecnologia (FCT), with reference 
UIDB/50021/2020 and PhD grant SFRH/BD/139473/2018.

\bibliographystyle{IEEEtran}

\bibliography{mybib}

\begin{thebibliography}{10}
\providecommand{\url}[1]{#1}
\csname url@samestyle\endcsname
\providecommand{\newblock}{\relax}
\providecommand{\bibinfo}[2]{#2}
\providecommand{\BIBentrySTDinterwordspacing}{\spaceskip=0pt\relax}
\providecommand{\BIBentryALTinterwordstretchfactor}{4}
\providecommand{\BIBentryALTinterwordspacing}{\spaceskip=\fontdimen2\font plus
\BIBentryALTinterwordstretchfactor\fontdimen3\font minus
  \fontdimen4\font\relax}
\providecommand{\BIBforeignlanguage}[2]{{%
\expandafter\ifx\csname l@#1\endcsname\relax
\typeout{** WARNING: IEEEtran.bst: No hyphenation pattern has been}%
\typeout{** loaded for the language `#1'. Using the pattern for}%
\typeout{** the default language instead.}%
\else
\language=\csname l@#1\endcsname
\fi
#2}}
\providecommand{\BIBdecl}{\relax}
\BIBdecl

\bibitem{dellwo2003relations}
V.~Dellwo, P.~Wagner, M.~Sol{\'e}, D.~Recasens, and J.~Romero, ``Relations
  between language rhythm and speech rate,'' 2003.

\bibitem{wagner2004introducing}
P.~S. Wagner and V.~Dellwo, ``Introducing yard (yet another rhythm
  determination) and re-introducing isochrony to rhythm research,'' in
  \emph{Speech Prosody 2004, International Conference}, 2004.

\bibitem{juliao2022can}
M.~Juliao, A.~Abad, and H.~Moniz, ``Can prosody transfer embeddings be used for
  prosody assessment?'' in \emph{Speech Prosody 2022, International
  Conference}, 2022in press.

\bibitem{ramus1999correlates}
F.~Ramus, M.~Nespor, and J.~Mehler, ``Correlates of linguistic rhythm in the
  speech signal,'' \emph{Cognition}, vol.~73, no.~3, pp. 265--292, 1999.

\bibitem{grabe2002durational}
E.~Grabe and E.~L. Low, ``Durational variability in speech and the rhythm class
  hypothesis,'' \emph{Papers in laboratory phonology}, vol.~7, no. 1982, pp.
  515--546, 2002.

\bibitem{arvaniti2012usefulness}
A.~Arvaniti, ``The usefulness of metrics in the quantification of speech
  rhythm,'' \emph{Journal of Phonetics}, vol.~40, no.~3, pp. 351--373, 2012.

\bibitem{tilsen2008low}
S.~Tilsen and K.~Johnson, ``Low-frequency fourier analysis of speech rhythm,''
  \emph{The Journal of the Acoustical Society of America}, vol. 124, no.~2, pp.
  EL34--EL39, 2008.

\bibitem{liss2010discriminating}
J.~M. Liss, S.~LeGendre, and A.~J. Lotto, ``Discriminating dysarthria type from
  envelope modulation spectra,'' 2010.

\bibitem{leong2014impaired}
V.~Leong and U.~Goswami, ``Impaired extraction of speech rhythm from temporal
  modulation patterns in speech in developmental dyslexia,'' \emph{Frontiers in
  human neuroscience}, vol.~8, p.~96, 2014.

\bibitem{vaysse2021automatic}
R.~Vaysse, J.~Farinas, C.~Ast{\'e}sano, and R.~Andr{\'e}-Obrecht, ``Automatic
  extraction of speech rhythm descriptors for speech intelligibility assessment
  in the context of head and neck cancers,'' in \emph{INTERSPEECH 2021}.\hskip
  1em plus 0.5em minus 0.4em\relax ISCA: International Speech and Communication
  Association, 2021.

\bibitem{qian2020unsupervised}
K.~Qian, Y.~Zhang, S.~Chang, M.~Hasegawa-Johnson, and D.~Cox, ``Unsupervised
  speech decomposition via triple information bottleneck,'' in
  \emph{International Conference on Machine Learning}.\hskip 1em plus 0.5em
  minus 0.4em\relax PMLR, 2020, pp. 7836--7846.

\bibitem{qian2021global}
K.~Qian, Y.~Zhang, S.~Chang, J.~Xiong, C.~Gan, D.~Cox, and M.~Hasegawa-Johnson,
  ``Global rhythm style transfer without text transcriptions,'' \emph{arXiv
  preprint arXiv:2106.08519}, 2021.

\bibitem{van2008visualizing}
L.~Van~der Maaten and G.~Hinton, ``Visualizing data using t-sne.''
  \emph{Journal of machine learning research}, vol.~9, no.~11, 2008.

\bibitem{eyben2015geneva}
F.~Eyben, K.~R. Scherer, B.~W. Schuller, J.~Sundberg, E.~Andr{\'e}, C.~Busso,
  L.~Y. Devillers, J.~Epps, P.~Laukka, S.~S. Narayanan \emph{et~al.}, ``The
  geneva minimalistic acoustic parameter set (gemaps) for voice research and
  affective computing,'' \emph{IEEE transactions on affective computing},
  vol.~7, no.~2, pp. 190--202, 2015.

\bibitem{rousseeuw1987silhouettes}
P.~J. Rousseeuw, ``Silhouettes: a graphical aid to the interpretation and
  validation of cluster analysis,'' \emph{Journal of computational and applied
  mathematics}, vol.~20, pp. 53--65, 1987.

\bibitem{thornton1998separability}
C.~Thornton, ``Separability is a learner’s best friend,'' in \emph{4th Neural
  Computation and Psychology Workshop, London, 9--11 April 1997}.\hskip 1em
  plus 0.5em minus 0.4em\relax Springer, 1998, pp. 40--46.

\bibitem{acevedo2019measuring}
A.~Acevedo, S.~Ciucci, M.~Kuo, C.~Dur{\'a}n, and C.~V. Cannistraci, ``Measuring
  group-separability in geometrical space for evaluation of pattern recognition
  and embedding algorithms,'' \emph{arXiv preprint arXiv:1912.12418}, 2019.

\bibitem{veaux2016superseded}
C.~Veaux, J.~Yamagishi, K.~MacDonald \emph{et~al.}, ``Superseded-cstr vctk
  corpus: English multi-speaker corpus for cstr voice cloning toolkit,'' 2016.

\end{thebibliography}


\end{document}